\documentclass[letterpaper,11pt]{emulateapj}
\usepackage{epsfig}
\usepackage{natbib}
\usepackage{graphics}
\usepackage{graphicx}
\usepackage{multirow}
\usepackage{amssymb}
\usepackage{color}
\usepackage{boxedminipage}

\def \kms {{\rm km/s}}

\shorttitle{[3.6] Photometry}
\shortauthors{Sorce et al.}

\begin{document}

\title{The mid-infrared Tully-Fisher relation: Spitzer Surface Photometry}

\author{Jenny G. Sorce$^{1}$}
\email{j.sorce@ipnl.in2p3.fr}
\author{H\'el\`ene M. Courtois$^{1,2}$}
\author{R. Brent Tully$^{2}$}
\affil{$^1$Universit\'e Claude Bernard Lyon 1, Institut de Physique Nucleaire, Lyon, France} 
\affil{$^2$Institute for Astronomy, University of Hawaii, 2680 Woodlawn Drive, HI 96822, USA}

\begin{abstract}


The availability of photometric imaging of several thousand galaxies with the {\it Spitzer Space Telescope} enables a mid-infrared calibration of the correlation between luminosity and rotation in spiral galaxies.  The most important advantage of the new calibration in the $3.6\mu$m band, IRAC ch.1, is photometric consistency across the entire sky.  Additional advantages are minimal obscuration,  observations of flux dominated by old stars, and sensitivity to low surface brightness levels due to favorable backgrounds.  
Through Spitzer cycle 7 roughly 3000 galaxies had been observed and images of these are available at the Spitzer archive. In cycle 8 a program called {\it Cosmic Flows with Spitzer} has been initiated that will increase by 1274 the available sample of spiral galaxies with inclinations greater than 45 degrees from face-on suitable for distance measurements.  This paper describes procedures based on the photometry package Archangel that are being employed to analyze both the archival and the new data in a uniform way. We give results for 235 galaxies, our calibrator sample for the Tully-Fisher relation.  Galaxy magnitudes are determined with uncertainties held below 0.05 mag for normal spiral systems.  A subsequent paper will describe the calibration of the [3.6] luminosity$-$rotation relation.

\end{abstract}



\section{Introduction}

The Cosmic Microwave Background (CMB) temperature dipole \citep{1996ApJ...473..576F} is usually interpreted as a motion of our Galaxy of over 600~\kms\ but the considerable majority of the posited motion is developed on large scales with origins that are still poorly understood.  Our overarching goal is to measure distances, hence parse departures from the mean Hubble expansion, on scales extending to 200~Mpc.  We are gathering distance measures from a multitude of methods and contributors within a program that we are calling {\it Cosmic Flows} (see Appendix).  Of particular importance for us are distances accrued from the correlation between the rotation rate of a galaxy and its luminosity, the Tully-Fisher Relation: TFR \citep{1977A&A....54..661T}.  There are methodologies that provide distance estimates that are individually more accurate but an abiding advantage of the TFR is applicability to a large fraction of all galaxies over a wide range of environments and distances.  There is the prospect of utilizing the TFR to obtain distances to several tens of thousands of galaxies out to redshift $z\sim0.05$ ($\sim$ 200 Mpc) .\\

As steps toward the accumulation of an appropriate data set of distances, our Cosmic Flows Large Program on the 100m Green Bank Telescope and complementary southern  observations on the Parkes Telescope \citep{2009AJ....138.1938C,2011MNRAS.414.2005C} are providing us with quality rotation information from HI line profiles, and the merging of our own and literature optical photometry \citep{2011MNRAS.415.1935C} provides the other element that has permitted a modern re-calibration of the TFR at an optical band \citep{2012ApJ...749...78T}.  The data accumulated in support of this program, including the new material to be discussed in this paper, are made available at EDD, the Extragalactic Distance Database, accessed online at http://edd.ifa.hawaii.edu \citep{2009AJ....138..323T}.\\

It has long been appreciated that photometry in the infrared may offer advantages because of reduced extinction and because infrared flux arises in large measure from old stars that should dominate the inventory of baryonic mass \citep{1979ApJ...229....1A}.  Photometry at $K_s$ band from the Two Micron All Sky Survey (2MASS) has been used with the TFR \citep{2002A&A...396..431K}.  However a major concern with observations in the infrared from the ground is high and variable sky foreground.  Much of the flux from galaxies lies in extended components with surface brightnesses that are well below the ground based sky level. Flux at the extremities of galaxies is lost and very low surface brightness galaxies are not even seen. \\

Observations from space removes the problem of the high contamination by the Earth's atmosphere.  We have initiated a sub-program that we call {\it Cosmic Flows with Spitzer} (CFS) with NASA's {\it Spitzer Space Telescope} \citep{2004ApJS..154....1W}.  Observations have begun in cycle 8 during the post-cryogenic period to obtain wide-field images of galaxies with IRAC, the InfraRed Array Camera, in Channel 1 (ch.1).  The present paper describes our reduction and photometry analysis procedures. We will discuss the transformation steps from raw Spitzer Post Basic Calibrated Data (PBCD) obtained from the Spitzer archive to the parameter needed for the TFR calibration: apparent [3.6] magnitudes.  The photometry is carried out with a Spitzer-adapted version of Archangel \citep{2007astro.ph..3646S,2012PASA...29..174S}. We will discuss the corrections to be made to apparent magnitudes and conclude with a discussion of uncertainties.  A subsequent paper will discuss the calibration of the TFR at $3.6 \mu$m.

\section{[3.6] Band Data}

Once cryogens were exhausted on {\it Spitzer Space Telescope} useful observations were restricted to IRAC ch.1 (centered at $3.55 \mu$m which we round to $3.6 \mu$m) and ch.2 ($4.5 \mu$m).   For the purposes of measuring distances with the TFR the two channels are highly redundant. Given a choice with the availability of finite observing resources between more galaxies in one band versus fewer galaxies in two bands we chose more galaxies in one band.  In our CFS program we concentrate on IRAC ch.1 observations in the $3.6 \mu$m window that give us magnitudes [3.6] in the AB system.  This window provides observations with minimal dust extinction \citep{1984ApJ...285...89D}. It lies at a minimum of the zodiacal background radiation \citep{1998EP&S...50..507O}.\\

\begin{figure}[h!]
\includegraphics[scale=0.5]{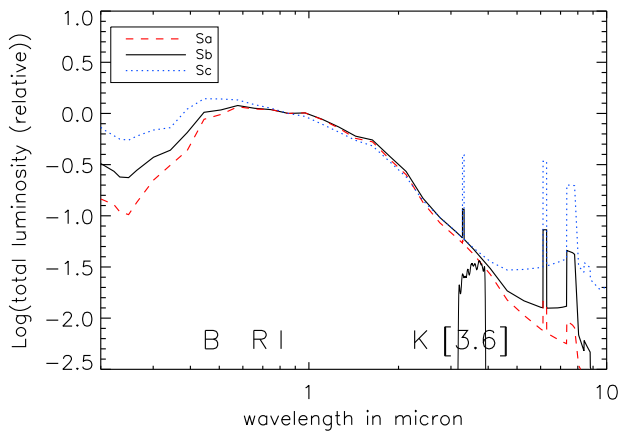}
\caption{Comparative SED for spirals of types Sa (red, dashed), Sb (black, solid), and Sc (blue, dotted). The Spitzer [3.6] passband is illustrated along with the wavelengths associated with $B,R,I,K$ bands.  The relative scales of the SED are offset to match at $0.8\mu$m.}
\label{sed}
\end{figure}

Figure~\ref{sed} provides examples of the spectral energy distribution (SED) of spiral galaxies \citep{1998ApJ...509..103S}.  The Spitzer [3.6] band lies on the Rayleigh-Jeans tail of the SED of normal populations of stars, not yet strongly affected by flux from warm dust that starts to become a factor at longer wavelengths than $4 \mu$m.  The discrete spectral features seen in the SED arise from Polycyclic Aromatic Hydrocarbon (PAH) molecules \citep{2008ARA&A..46..289T}.  The highest frequency PAH, at $3.3 \mu$m is contained within the [3.6] bandpass.  \citet{2012ApJ...744...17M} have investigated the impact of various contributors to flux in the $3.6 \mu$m window with 6 representative spiral galaxies observed with the Spitzer S4G program.  They find contributions from hot dust and PAHs together contribute $9\pm4$\% of the global flux in the $3.6 \mu$m band, intermediate age asymptotic giant branch and red supergiant branch stars contribute $3\pm2$\% of the global flux, and the rest, the great majority, is contributed by old stars, predominantly K and M giants.  These non-stellar and young stellar contributions should only slightly degrade the correlation between old stars and mass in normal spirals.\\ 

Using Spitzer IRAC ch.1, a point spread function with mean FWHM $1.66^{\prime\prime}$ is sampled with $1.2^{\prime\prime}$ pixels.  The field of view is $5.2^{\prime}$, adequate to encompass most galaxies to beyond twice $d_{25}$, the diameter at a $B$ isophote of 25 mag per square arcsec.  Larger galaxies require mosaics.  Integrations with the CFS program involve the combination of 8 x 30 second slightly dithered exposures for a total of 4 minutes per field.  As will be discussed, these integrations provide images that probe somewhat fainter limits than most ground-based optical photometry programs and much fainter limits than ground-based infrared photometry programs.
Spitzer surface brightness levels reach ten magnitudes below typical ground-based infrared sky levels.  No existing near-infrared ground survey achieves the accuracy obtained with Spitzer Space Telescope.   The outstanding advantages of space observations are background stability and all-sky consistency \citep{2004ApJS..154...10F}.\\

Our cycle 8 post-cryogenic program CFS avoids repetition of earlier Spitzer observations.  Archival information is used where available.  Major contributions from earlier programs come from SINGS, the Spitzer Infrared Nearby Galaxies Survey \citep{2005ApJ...633..857D,2007ApJ...655..863D} and LVL, the Local Volume Legacy survey \citep{2009ApJ...703..517D} carried out during the cryogenic phase, then $S^{4}G$, the Spitzer Survey of Stellar Structure in Galaxies \citep{2010PASP..122.1397S}, and CHP, the Carnegie Hubble Program \citep{2011AJ....142..192F}, subsequently carried out during the post cryogenic phase. Smaller programs undertaken during the cryogenic phase supply us with a few more fields. These data are available for public use at the Spitzer Heritage Archive website (SHA\footnote{http://irsa.ipac.caltech.edu/data/SPITZER/docs/spitzerdataarchives/}).  The variety of source programs introduces variations in the details of the acquisition, particularly affecting total integrations, dithering procedures, and the extent of fields referenced to $d_{25}$.  However, with all the data that will be considered the fields are large enough and the exposure times are long enough that at most only a few percent of the light from a target is lost.


\section{Photometry}

\subsection{Surface Photometry}

Large numbers of pixels complicate simple parameter extractions. A galaxy is spread over a large area of the sky. At some point outer pixels have more sky luminosity (zodiacal light and background contaminants) than galaxy luminosity. Setting the "sky" dominates the total magnitude error budget. An analysis of a large galaxy (extending across many pixels) requires surface photometry involving fits of isophotes, lines of constant luminosity. Isophotes are often set to be ellipses \citep{1999BaltA...8..535M}. Our interest is with spiral galaxies with types typically between Sa and Scd.  A well behaved spiral is approximated by an oblate spheroid that appears circular when viewed face-on and projects to an ellipse when viewed toward edge-on. 
Galaxy 2D images described by elliptical isophotes can be summed in annuli to reduce to a 1D description. Then, the 1D profiles are fitted by various functions in order to extract the radial surface brightness (SB) distribution, global structure or geometrical characteristics, spatial orientation, stellar populations, characteristics of dust, etc. To obtain apparent magnitudes, \citet{1977ApJS...33..211D} introduced the growth curve, a plot of magnitude within a radius as a function of radius. With an adequate signal to noise ratio, it could be enough to place large apertures around galaxies and sum the total amount of light, minus the sky contribution. In practice, a galaxy luminosity distribution decreases towards larger radii so larger apertures catch more galaxy light but also introduce more sky noise. Some light is inevitably lost below the sky level. Isophotal intensities associated with the galaxy light at large radii are sensitive to the sky setting. Restriction to a smaller radius leads to underestimates of total light. The problem is that galaxies do not have discrete edges.\\

It is never possible to measure 100\% of the light of a galaxy. Measurements are made to an isophotal level dictated by telescope optics, detector, exposure times, and sky brightness. Different authors measure magnitudes to different isophotal levels then often extrapolate to total magnitudes. Our interest is with spiral galaxies. These galaxies characteristically decay exponentially in luminosity with radius. In an ideal case, light contained within a specified isophotal level is a simple function of the disk central SB and the exponential decay scale length.  To extrapolate in such a case one can assume that the light at large radii falls off like an exponential disk with a central SB and scale length characterized by a fit to the main body of the galaxy. The estimated contribution lost below the sky level can be added to what is observed to give an extrapolated magnitude \citep{1996AJ....112.2471T}.

\subsection{Archangel}

\begin{figure}[h!]
\includegraphics[scale=0.37]{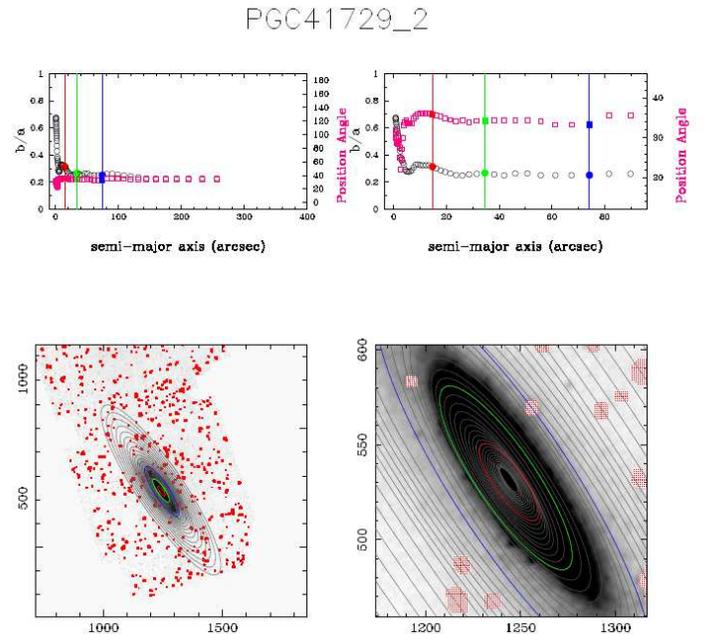}
\caption{Output of the Archangel software showing the ratio b/a and the position angle in the two top panels and the fitted ellipses and the masking in the two bottom panels for PGC41729 = NGC4522.}
\label{ellipses}
\end{figure}

	\citet{2007astro.ph..3646S} developed Archangel, a flexible tool for galaxy surface photometry built of a combination of FORTRAN and Python routines. Archangel performs procedures such as: 1) masking of stars and flaws, 2) ellipse fitting at expanding radii from the galaxy center, 3) compression of 2D information into 1D SB and magnitude growth curves as a function of radius, and 4) extrapolation via fits to the magnitude growth curve at large radii involving rational functions \citep{2012PASA...29..174S}.  Position angles and ellipticities are freely determined at each radial step in the development of the growth curve.  At large radii noise dominates and position angle and ellipticity are frozen for the remaining outward steps in radius.  The program provides flexibility in where these parameters become frozen and allows that they may be frozen at all radii.  Total magnitudes, the most important product of this analysis, are found to be negligibly affected by position angle and ellipticity details at intermediate radii.  Comparisons with alternative photometry are discussed in section 5.1.  See Figure~\ref{ellipses} for an example of masking and ellipse fitting with Archangel.\\

To obtain magnitudes in the AB system from the archival PBCD we use magnitude and VEGA/AB conversion parameters from the IRAC Instrument handbook\footnote{http://irsa.ipac.caltech.edu/data/SPITZER/docs/irac/ iracinstrumenthandbook/} and from \citet{2006MNRAS.366..609C} respectively.  At optical bands it is common practice to quote magnitudes in the Vega photometric system but working in the mid-infrared it is more useful to use the AB system.  In instances where comparisons are made between optical and mid-infrared, we use the following transformations \citep{1994AJ....108.1476F}:
$$B(Vega) = B(AB) + 0.163$$
$$R_{C}(Vega) = R_{C}(AB) - 0.117$$
$$I_{C}(Vega) = I_{C}(AB) - 0.342$$
$$[3.6](Vega) = [3.6](AB) - 2.785$$

	A significant source of uncertainty arises from the setting of the sky level (see section 5). In Archangel the sky is taken as the median of sky boxes placed around the galaxy. This method gives realistic initial sky background estimates \citep{2011arXiv1111.5009H}. If targets are modest in size there is reasonable control of the sky level. If the sky is set properly then the magnitude growth curve should go asymptotically flat at large radii.  One can also evaluate the sky setting by looking at the SB as a function of radius. SBs are not expected to flare or drop precipitously at the sky level, although such occurrences are not phenomenologically excluded \citep{2003ApJ...582..689M, 2008AJ....135...20E}. Visual inspections of the magnitude growth curve and SB dependence with radius ensure an optimal sky setting. Fortunately, sky values are low in Spitzer data even if we will show in the last section that this problem remains our major source of uncertainty.\\
	
An issue related to the sky problem is the matter of the terminal radius of an analysis.   A limit to the fitting process can be imposed by signal-to-noise considerations. Integration times permit us to reliably reach a radius $a_{26.5}$ at the isophotal level 26.5~mag~arcsec$^{-2}$ in the [3.6] band.  We try to extend the ellipse fitting to $1.5 a_{26.5}$ in the [3.6] band.  A goal of the program is to assure that the ellipse fitting extends to at least $1.1 a_{26.5}$, with mosaics if necessary.
This [3.6] band dimension is not available before the observation so we rely on a substitution found to be comparable based on the $B$ band diameter $d_{25}$, requiring that the observed area extend to a radius $1.5 d_{25}$ \citep{2010PASP..122.1397S}.

\subsection{1D Fits and Parameter Extractions}

The mean SB in magnitudes per square arcsec in an annulus at radius $r_i$ depends on the mean flux in a pixel at that radius $F(r_i)$ and the mean sky flux in a pixel $S$:

\begin{equation}
\mu(r_i) = -2.5 {\rm log} (\frac{F(r_i)-S}{0.6}) + 21.585
\label{SB_equation}
\end{equation}
where the constant in the denominator provides a conversion from pixels to arcsec.\\


\begin{figure}[h!]
\includegraphics[scale=0.48]{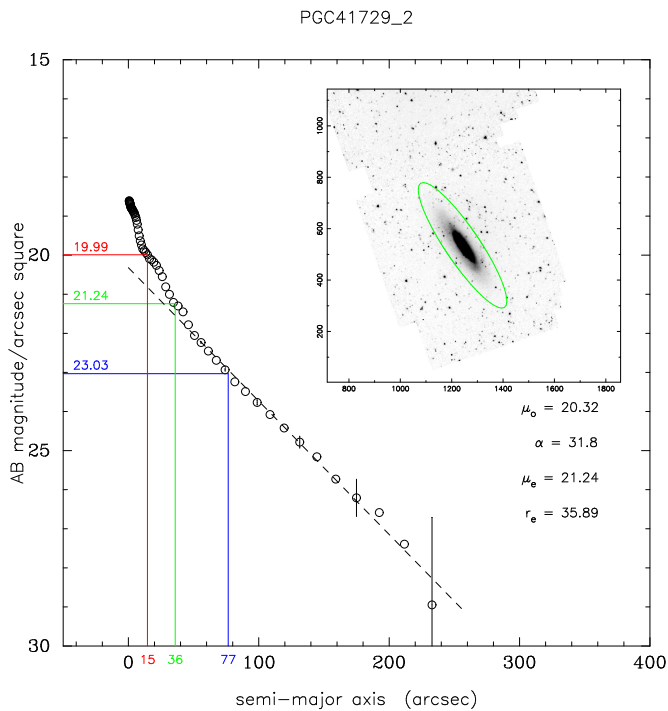}
\caption{Output of the Archangel software showing the SB profile for PGC41729=NGC4522.   The red, green, and blue lines identify the radii enclosing 20\%, 50\%, and 80\% of the total light respectively and the surface brightnesses at those radii. The dashed straight line illustrates an exponential fit to the disk between the radius enclosing 50\% of the light and the isophotal radius $a_{26.5}$. The insert provides a [3.6] image of the galaxy with an annulus at the isophote 26.5~mag~arcsec$^{-2}$.}
\label{SB}
\end{figure}

Archangel allows for a description of the run of SB with radius as the sum of disk and bulge components.  Instead, we choose to restrict to disk fits only.   With multiple component fits there are frequently trade-offs such that the overall fit may be satisfactory but the physical meanings of parameters are ambiguous.  Usually the dominant radial SB characteristic of spiral galaxies is an exponential decay of projected luminosity with radius.  Deviations are most frequently seen toward the center where a bulge may become dominant. It is beyond the scope of this program to dissect galaxy images into detailed morphological components because such dissection has negligible effect on the product that most interests us: total magnitudes.  We restrict fitting to a rough characterization of the exponential fall-off.\\

The radial run of SB in an exponential disk, $\mu_d(r)$,  has a behavior described by an e-folding scale length, $\alpha$, and central SB, $\mu_0$ 
\begin{equation}
\mu_d(r) = \mu_0 +  1.0857 (r/\alpha).
\label{disk}
\end{equation}
As a procedure, we determine the disk parameters with a fit over the range of radii from the half-light radius $a_e$ to the isophotal limit radius $a_{26.5}$.  This range is modified if a visual inspection indicates the need.  An example of a SB product is illustrated in Figure~\ref{SB}.\\

\begin{figure}[h!]
\includegraphics[scale=0.48]{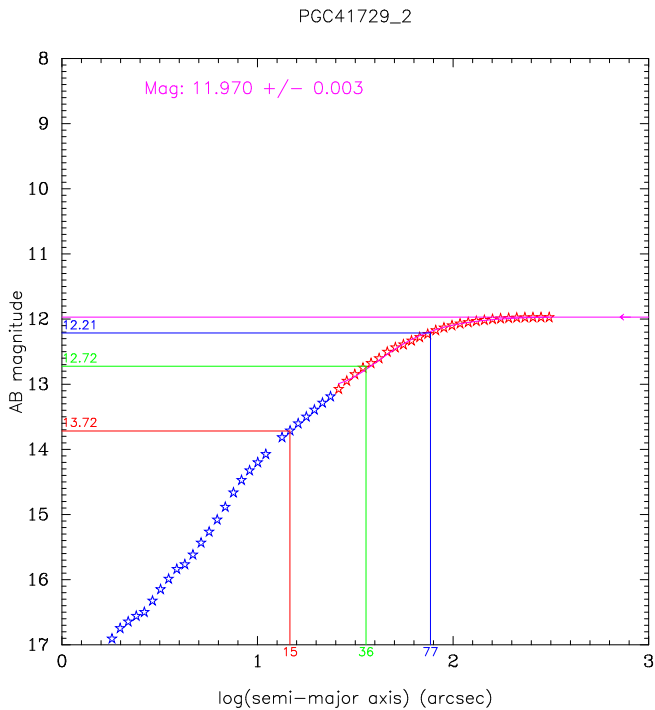}
\caption{Output of the Archangel software showing the growth curve for PGC41729=NGC4522.  The incremental growth of the apparent magnitude of the galaxy with radius is shown by the progression of stars.  The fit providing an extrapolation to a total magnitude is generated over the domain of the red stars.  The level of the total magnitude is shown in magenta.  The 80\%, 50\%, and 20\% light enclosure levels are shown in blue, green, and red.}
\label{growthcurve}
\end{figure}

An example of a magnitude growth curve as a function of semi-major axis is shown in Figure~\ref{growthcurve}.  The light from each succeeding annulus contributes to the (negatively) increasing magnitude with increasing radius.  If the sky value is properly set then the growth curve will asymptotically flatten.  Should the curve turn over it would be inferred that the sky level is set too high - flux from the galaxy is being attributed to the sky and being removed.  Conversely, the sky set too low causes flux from the sky to be attributed to the galaxy and the growth curve will fail to flatten.\\

The Spitzer photometry is sufficiently deep that magnitudes in the growth curve approach the total magnitude of the galaxy.  One way to extend to the total magnitude uses the procedure built into Archangel based on interpolations and extrapolations with rational functions. 
Such functions have a wide range in shape and have better interpolating properties than polynomial functions. They suit data where an asymptotic behavior is expected. The quadratic form/quadratic form, meaning a degree of 2 in both numerator and denominator is the simplest choice. The asymptotic magnitude is $c_1/c_2$ where $c_1$ and $c_2$ are the second order coefficients of the numerator and denominator respectively. However, rational functions are non-linear. They can produce vertical asymptotes due to roots in the denominator that are to be ignored.  Fit uncertainties are given by the standard error of the estimate, SEE: 
\begin{equation}
SEE = \sqrt{\frac{1}{n} \sum_{i=1}^{n} (m(a_i)_{fit} - m(a_i)_{measured})^{2}}.
\label{error}
\end{equation}
 
Given a growth curve as seen in Figure~\ref{growthcurve} it is straightforward to define the useful parameters $a_{20}$, $a_e$, and $a_{80}$ enclosing 20\%, 50\%, and 80\% of the light respectively.   The associated magnitudes and semi-major radii are illustrated in Figures 2$-$4.  Other products are the average SB within $a_e$ and $a_{20}$ and a concentration index $C_{82} = a_{80}/a_{20}$. Table~\ref{table1} gives the parameters that are extracted for the galaxy used as an example in Figs. 2$-$4 and illustrates what is seen in a single row in the catalog {\it Spitzer [3.6] Band Photometry} at EDD, the Extragalactic Distance Database\footnote{http://edd.ifa.hawaii.edu}. Capabilities within EDD allow a user to link to other catalogs, thereby accessing all manner of information about each target.\\

We provide three magnitudes that approximate the global magnitude of the galaxy: $[3.6]_{26.5}$, an isophotal magnitude that directly measures the light within a reliably attained radius, $[3.6]_{tot}$, a `total' magnitude given by the asymptote of the Archangel rational function extrapolation, and $[3.6]_{ext}$, an `extrapolated' magnitude assuming a continuation of the exponential disk beyond the radius of the isophote 26.5 mag~arcsec$^{-2}$.  The relative merits of these magnitudes will be discussed in a later section. 



\begin{table}[h!]
\begin{center}
\begin{tabular}{|c|c|c|c|}
\hline
1 & 2 & 3 & 4 \\
\hline
PGC & Name & Date & Exp  \\
\hline
41729 & NGC4522 & 2007.02.14T14:46:36.378 & 240  \\
\hline
\end{tabular}
\end{center}
\end{table}

\vspace{-0.8cm}

\begin{table}[h!]
\begin{center}
\begin{tabular}{|c|c|c|c|c|c|c|}
\hline
5 & 6 & 7 & 8 & 9 & 10 & 11 \\
\hline
$a_{26.5}$ & $[3.6]_{26.5}$ & $[3.6]_{tot}$ & $\sigma_{m}$ & $[3.6]_{ext}$ & $\mu_0$ & $\alpha$ \\
\hline
181 & 11.98 & 11.970 & 0.003 & 11.957 & 20.32 & 31.8 \\
\hline
\end{tabular}
\end{center}
\end{table}

\vspace{-0.8cm}

\begin{table}[h!]
\begin{center}
\begin{tabular}{|c|c|c|c|c|c|c|c|}
\hline
12 & 13 & 14 & 15 & 16 & 17 & 18 & 19 \\
\hline
b/a & $\sigma_{b/a}$ & PA & $a_{80}$ & $\mu_{80}$ & $a_e$ & $\mu_e$ & $<\mu_e>$ \\
\hline
0.26 & 0.01& 34 & 77 & 23.04 & 36 & 21.24 & 20.31 \\
\hline
\end{tabular}
\end{center}
\end{table}

\vspace{-0.8cm}

\begin{table}[h!]
\begin{center}
\begin{tabular}{|c|c|c|c|c|}
\hline
20 & 21 & 22 & 23 & 24 \\
\hline
 $a_{20}$ & $\mu_{20}$ & $<\mu_{20}>$ & $C_{82}$  & RefLink \\
\hline
15 & 19.99 & 19.53 & 5.2 & SSOV \\
\hline
\end{tabular}
\caption{Extracted photometry parameters. (1) Principal Galaxies Catalog number, (2) common name, (3) date of Spitzer observation, (4) nominal total integration, seconds (actual time collecting photons somewhat less), (5) $a_{26.5}$: major axis radius at isophote 26.5 mag arcsec$^{-2}$, (6) $[3.6]_{26.5}$: AB magnitude within $a_{26.5}$, (7) $[3.6]_{tot}$: total AB magnitude from rational function asymptote, (8) $\sigma_m$: rms deviations, rational function fit, (9) $[3.6]_{ext}$: total AB magnitude by extrapolating flux beyond $a_{26.5}$ assuming continuance of exponential disk, (10) $\mu_0$: central disk surface brightness from inward extrapolation of disk fit, mag arcsec$^{-2}$, (11) $\alpha$: exponential disk scale length, arcsec, (12) $b/a$: ratio of minor to major axes, (13) $\sigma_{b/a}$: uncertainty in axial ratio, (14) PA: position angle of major axis, deg. (15) $a_{80}$: major axis radius of annulus enclosing 80\% of total light, arcsec, (16) $\mu_{80}$: surface brightness at $a_{80}$, mag arcsec$^{-2}$, (17) $a_e$: `effective radius', major axis radius of annulus enclosing 50\% of total light, arcsec, (18) $\mu_e$: surface brightness at $a_e$, mag arcsec$^{-2}$, (19) $<\mu_e>$: average surface brightness within $a_e$, mag arcsec$^{-2}$, (20) $a_{20}$: major axis radius of annulus enclosing 20\% of total light, arcsec, (21) $\mu_{20}$: surface brightness at $a_{20}$, mag arcsec$^{-2}$, (22) $<\mu_{20}>$: average surface brightness within $a_{20}$, mag arcsec$^{-2}$, (23) $C_{82}$: concentration index, $a_{80}/a_{20}$, (24) Spitzer program link.} 
\label{table1}
\end{center}
\end{table}






\section{Corrections}

Corrected apparent magnitudes $[3.6]^{b,k,i,a}$ for Spitzer IRAC ch.1 data are given by:
\begin{equation}
[3.6]^{b,i,k,a}= [3.6] - A_b^{[3.6]} - A_i^{[3.6]} - A_k^{[3.6]}+A_a^{[3.6]}
\label{corrected_mag}
\end{equation}
with apparent magnitude [3.6] output from Archangel, galactic extinction correction $A_b^{[3.6]}$, internal extinction correction $ A_i^{[3.6]}$, $k$-correction $A_k^{[3.6]}$, and aperture correction $A_a^{[3.6]}$. We shall describe these terms in the following sub-sections.

\subsection{Galactic extinction correction}

 	Galactic extinction depends only on object coordinates and observational wavelengths. The InfraRed Science Archive (IRSA) provides an online tool at http://irsa.ipac.caltech.edu/applications/DUST/ with $100 \mu$m cirrus maps \citep{1998ApJ...500..525S} that supplies us with differential reddenings, $E(B-V)$. We use the correction term given by \citet{1989ApJ...345..245C} accounting for a small shift to the centroid of the Spitzer passband:
\begin{equation}
A_{b}^{[3.6]} = R_{[3.6]} E(B-V)
\end{equation}  
with $R_{[3.6]} = 0.20$.  Galactic extinction magnitude corrections at [3.6] are only 9\% compared to at $I_{C}$ and 4\% of the corrections at $B$. Corrections at latitudes above $15^{\circ}$ are almost always 0.05 mag or less, with uncertainties $\sim 0.01$ mag.\\

\subsection{Internal extinction correction}

Internal extinction is usually the greatest concern. Fortunately, in the infrared such extinction is very small. \citet{1995AJ....110.1059G,1997AJ....113...22G} showed that there is a luminosity dependence to galaxy internal obscurations. \citet{1998AJ....115.2264T} confirmed and provided an alternative description of the effect. There is a subtle problem because absolute magnitudes are not known a priori.  They are a product of the analysis. \citet{1998AJ....115.2264T} framed magnitude corrections in term of a distance-independent surrogate, the line width parameter, $W_{mx}^i$.  Accordingly, the internal extinction correction can be written:
\begin{equation}
 A_i^{[3.6]}=\gamma_{[3.6]} {\rm log}(a/b)
 \end{equation}
where $\gamma_{[3.6]}$ is:
\begin{equation}
\gamma_{[3.6]} = 0.10 + 0.19({\rm log}~W^{i}_{mx} - 2.5)
\end{equation}  
if $W^i_{mx} > 94$~km s$^{-1}$ and $\gamma_{[3.6]}=0$ otherwise.
$W_{mx}^i$ is a measure of twice the maximum rotation rate of a galaxy derived from $W_{m50}$, the HI profile width at $50\%$ of the mean flux within the velocity range encompassing $90\%$ of the total HI flux \citep{2009AJ....138.1938C,2011MNRAS.414.2005C}. The measure includes a deprojection appropriate for the inclination $i$ (see the above references for the derivation of inclinations).\\

There is an advantage to this formulation of the internal extinction. If the inclination is underestimated, ${\rm log} (a/b)$ is underestimated driving $A^{[3.6]}_i$ low but then $W_{mx}^{i}$ is overestimated which drives $\gamma_{[3.6]}$, hence $A^{[3.6]}_i$ up. The two terms in $A_i^{[3.6]}$ are affected in opposite directions. Regardless, internal absorption corrections are always small, rarely reaching 0.1 mag.  Uncertainties in these corrections are less than 0.02 mag.\\

\subsection{K-correction}

	The $k$-correction \citep{1968ApJ...154...21O} is small at the redshifts we encounter. \citet{2007ApJ...664..840H} show a linear dependence of the $k$-correction with redshift at $3.6 \mu$m. This linear dependence is independent of the galaxy type at small redshifts at this position on the Rayleigh-Jeans tail of the spectral energy distribution of star light.
	We use the low-$z$ formulation by \citet{2007ApJ...664..840H}:
\begin{equation}
A_k^{[3.6]} = -2.27 ~z
\end{equation}
with $z$ the galaxy redshift.  Uncertainties are at the level of 0.01 mag or less.\\

\subsection{Aperture correction} 

	The fourth and last adjustment is the aperture correction. Aperture corrections are required for extended source photometry with Spitzer (e.g. galaxies) because their absolute calibrations are tied to point sources with IRAC observations. There is extended emission from the Point Spread Function outer wings, and the scattering of the diffuse emission across the focal plane that is captured by the extended source photometry but not by the calibrations on point sources. Since the photometry is normalized to 12" apertures a correction must be applied for large apertures (\citet{2005PASP..117..978R} and IRAC Instrument Handbook). The following correction is recommended\textbf{\footnote{http://irsa.ipac.caltech.edu/data/SPITZER/docs/irac/ iracinstrumenthandbook/}}. For an effective aperture radius $r$ in arcseconds, the ch.1 IRAC extended source aperture correction is $f_{IRAC\,true} = f_{IRAC\,measured}  \times (Ae^{-r^B} + C)$ where $A=0.82$, $B=0.37$ and $C=0.91$. The extended source aperture correction in magnitudes is:\\
\begin{equation}
A_a^{[3.6]}=-2.5{\rm log}(Ae^{-r^B} + C).
\end{equation}
The average correction for galaxies of interest to our program is 0.10.
The variations on this correction from source to source for our galaxies, which are typically larger than $1^{\prime}$, is 0.01 mag and 10\% relative uncertainties in the adjustment are negligible.


\section{Uncertainties}

An extremely important virtue of the Spitzer [3.6] band photometry is the robustness of the luminosity measurements (a) with uniformity across the sky, (b) with inclusiveness of target light because of the sensitivity, and (c) because adjustments are small.  There was a discussion of uncertainties associated with the adjustments in the last section and it can be summarized that as long as sources are not in extremely obscured regions of our Galaxy ($A_b< 1$) then the global {\it uncertainty} in adjustments is at the level of 0.03 mag or less, with internal absorption within sources dominant in the error budget.  The IRAC handbook gives a $2-3\%$ error on the absolute flux calibration (excluding the aperture correction), but more importantly for this program, claims photometry is repeatable across the sky at the 1\% level.\\

Among our parameters we determine isophotal, `total', and `extrapolated' magnitudes. The latter two both approximate the global magnitude, the `total' from the rational function asymptote of the growth curve and the `extrapolated' from the extension of the exponential disk fit beyond the radius of the isophotal magnitude. By construction, $[3.6]_{26.5}$ is fainter than $[3.6]_{ext}$ and should be fainter than $[3.6]_{tot}$.  The average difference $<[3.6]_{26.5} - [3.6]_{ext}> = 0.016$ mag corresponds to a typical disk fit of 6.2 exponential scalelengths at the 26.5 mag/arcsec$^{-2}$ isophote.  The typical {\it uncertainty} in this extrapolation is below 0.01 mag except if the target is extremely low surface brightness.  SB profiles of spirals can depart from a pure exponential at large radii, either with flares or truncations and because of the interplay between bulges and disks \citep{1985ApJS...59..115K}.  Yet because such a large fraction of the flux is captured by the deep Spitzer integrations the differences between measured isophotal and extrapolated magnitudes are so small as to leave little room for uncertainty in the extrapolation. \\

By comparison, $<[3.6]_{26.5} - [3.6]_{tot}> = 0.007$ mag, that is, $[3.6]_{tot}$ is fainter than $[3.6]_{ext}$ by 0.009 mag on average. The rms scatter is 0.018 mag between these alternative measures.  The differences are primarily due to a slight instability in the rational function fits.  We give preference to the \emph{exponential disk extrapolations}. \\

We turn to what is probably the largest source of error, the setting of the `sky' level.  With observations in space at [3.6] band this noise level is dominated by diffuse zodiacal light and discrete high redshift galaxies.  The discrete contaminants can be easily seen to very faint levels in regions beyond the galaxy.  They are less easy to see and exclude if they are superimposed on the target galaxy.   A major task before running a surface photometry analysis is the removal of contaminants like foreground stars and background galaxies.  Our approach is to not be too aggressive with the removal of contaminants.  We remove contaminants as best we can on the target and remove contaminants in the adjacent sky to the same level, leaving in place fainter sources since such sources must also be hidden within the galaxy.\\

It was described in the section on Archangel photometry that sky settings were established from the median of pixel fluxes in boxes placed around the galaxy and validated by the nature of the magnitude growth curve (it should go asymptotically flat) and the surface brightness profiles (flares or cutoffs as noise dominance is approached are suspicious but not considered a conclusive sign of bad sky setting).  In order to generate a quantitative test of the effects of sky variance we have run Spitzer-adapted Archangel on 235 galaxies, our calibrator sample defined in \citet{2012ApJ...749...78T}. Our calibrator sample contains only a few LSB and irregular galaxies.\\

\begin{figure}[h!]
\includegraphics[scale=0.52]{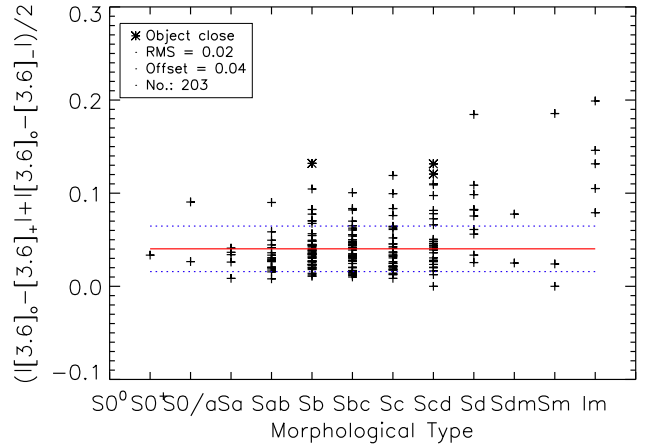}
\caption{Variation of magnitude uncertainty as a function of morphological type.  The mean offset of 0.04 mag and rms scatter of 0.02 mag indicated by the solid red and dotted blue lines respectively excludes types Sd and later.  Three cases with contamination from nearby bright objects are indicated by asterisks. The scatter is asymmetric about the mean since an absolute value difference from the fiducial value cannot be less than zero.}
\label{type}
\end{figure}

\begin{figure}[h!]
\includegraphics[scale=0.52]{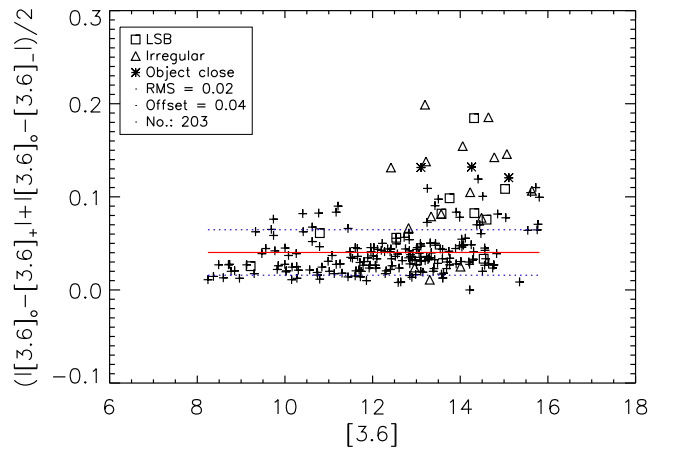}
\caption{Variation of magnitude uncertainty as a function of magnitude. Type Sd systems, here referred to as low surface brightness galaxies, are represented by squares and types Sdm-Sm-Im irregular galaxies are represented by triangles. Asterisks locate galaxies with a very bright object close to them. The mean offset and scatter lines have the same meaning as in the previous figure.}
\label{mag}
\end{figure}

A first run gives us the sky value $S_0$ and its uncertainty. We run Archangel two more times with sky values of $S_0\pm\sigma_{{\rm sky}}$ respectively for each one of our selected galaxies. This gives us three extrapolated magnitudes that we call [3.6]$_0$, [3.6]$_+$, and [3.6]$_-$. Figures \ref{type} and \ref{mag} show the variation of $(|[3.6]_0-[3.6]_-|+|[3.6]_0-[3.6]_+|)/2$ as functions of type and apparent magnitude. These plots show the sensitivity to the choice of sky value and that this sensitivity becomes particularly acute for low surface brightness systems.  The galaxies of type Sd, identified in the plots as low surface brightness galaxies (LSB), and the magellanic irregular galaxies are clear outliers. There are also 3 galaxies with very bright objects nearby that could influence the sky level with stray light and might explain their position in Figure \ref{mag}. These 3 are retained in the calculations of the offset and scatter since they are typical spiral galaxies but the LSB and irregular galaxies are excluded.  Magnitude uncertainties due to the sky error are of the order $0.04\pm0.02$.   Uncertainties with LSB and irregular galaxy magnitudes tend to be more important which is understandable. LSB have surface brightness values closer to that of the sky and irregular galaxies not only tend to be LSB but in addition might not be well described by elliptical isophotes. Changing the sky value a little might change the measured flux considerably toward the external part of such galaxies.  LSB and irregular galaxies apart, Figure \ref{mag} indicates that uncertainties do not strongly increase at fainter magnitudes.\\

\begin{figure}[h!]
\includegraphics[scale=0.52]{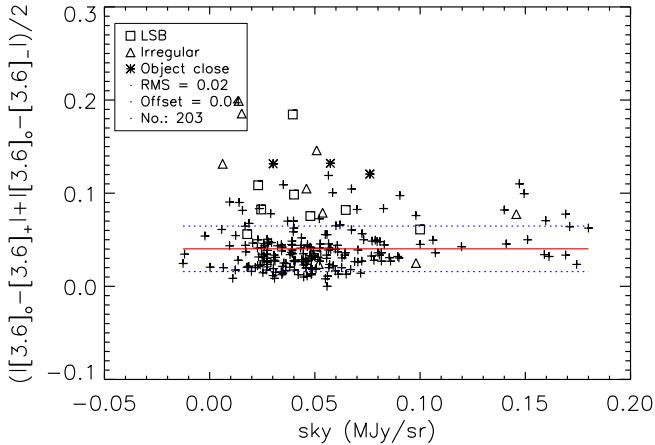}
\caption{Variation of magnitude uncertainty as a function of the sky value in MJy~sr$^{-1}$.  Squares represent low surface brightness galaxies while triangles stand for irregular ones. Galaxies represented by asterisks are galaxies with a very bright object close to them. Sky uncertainties can be 4 times higher than normal without resulting in abnormally high uncertainty in magnitudes.}
\label{sky}
\end{figure}

\begin{figure}[h!]
\includegraphics[scale=0.52]{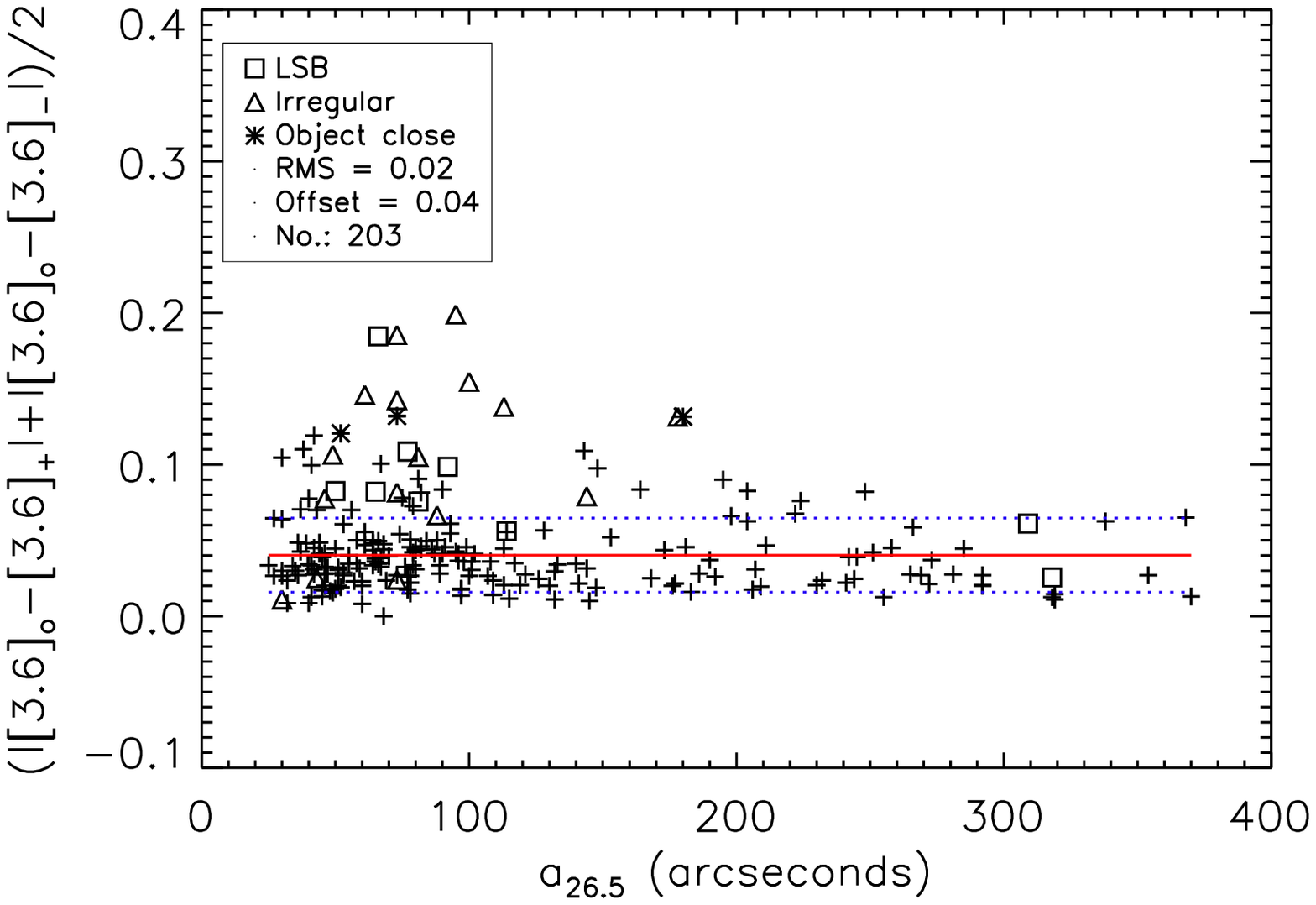}
\includegraphics[scale=0.52]{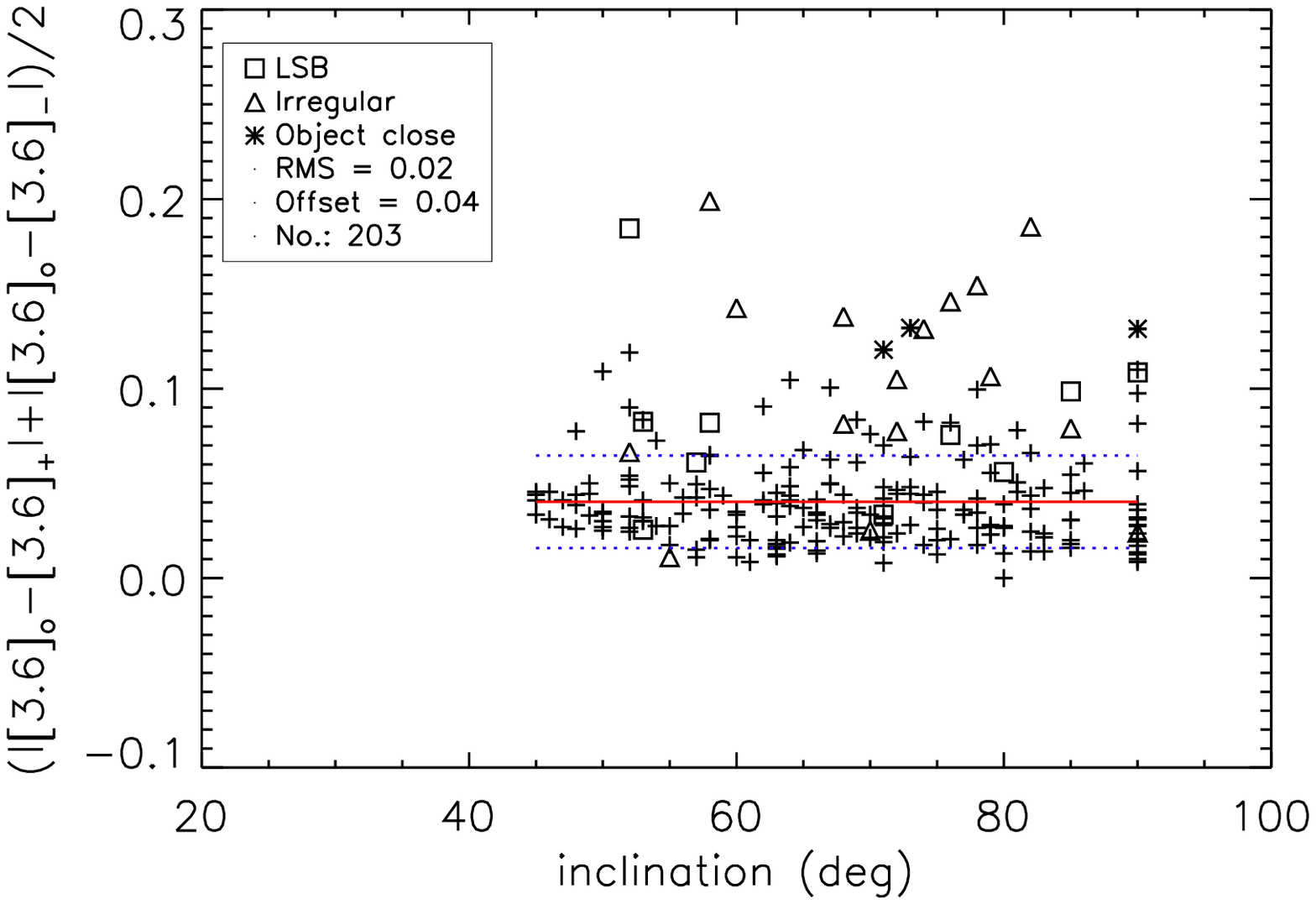}
\includegraphics[scale=0.52]{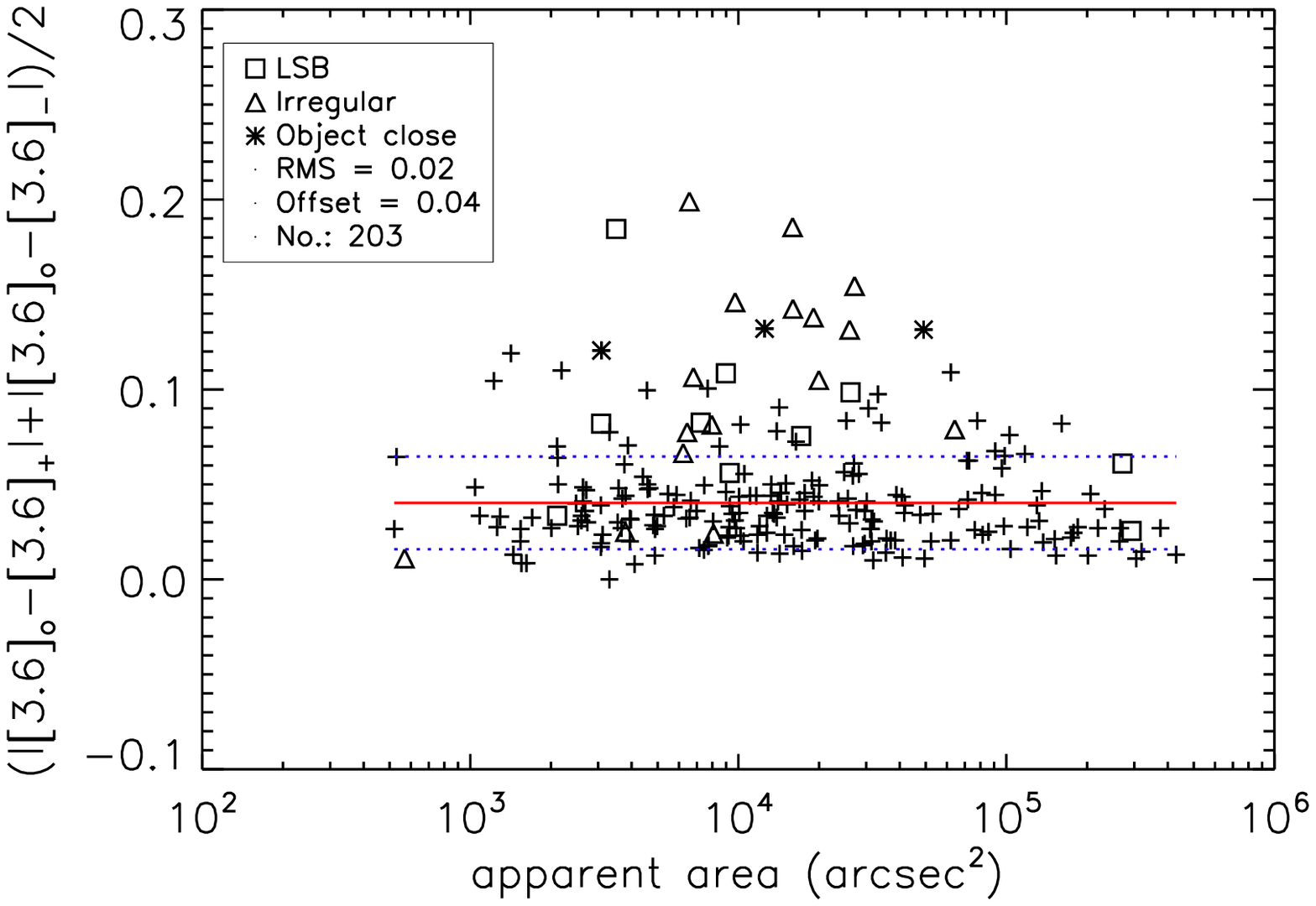}
\caption{Variation of magnitude uncertainty as a function of the galaxy: {\it (top)} semi-major axis, {\it (middle)} inclination, and {\it (bottom)} apparent area. Squares, triangles and asterisks represents low surface brightness, irregular, and possibly contaminated galaxies respectively. There is no apparent correlation between magnitude uncertainty and the radius, inclination, or apparent area of a galaxy.}
\label{allother}
\end{figure}

Next we test for uncertainties in magnitude due to the sky against the sky value itself, as well as against the isophotal semi-major axis in the [3.6] band $a_{26.5}$, inclination from face-on, and apparent area defined as the area of the ellipse at $a_{26.5}$ to see if any trends exist.  The results in Figures \ref{sky} and \ref{allother} show no correlation. In Figure \ref{sky}, we can see that the uncertainty in magnitude does not depend on the sky value. We checked for a dependence on sky uncertainty and find no correlation. These results suggest that the total and extrapolated disk apparent magnitudes are adequate (we do not show the plots for both magnitudes here as they are very similar).  In any case, the highest sky values are relatively moderate ($< 0.20$ MJy sr$^{-1}$). One can also notice that sky values and sky uncertainties are not correlated, evidently a reflection of the relative uniformity of background across dimensions of $5-10$ arcmin.  Structure in the background could be a worse problem when the sky setting is very low. Perhaps it is a surprise that the uncertainty is not proportional on the galaxy apparent area (Figure \ref{allother}, bottom). The more pixels that are affected by setting a new sky, the more the magnitude might change. In any case, these tests indicate that magnitude uncertainties can be taken to be approximately constant for all normal spiral galaxies.

\subsection{Comparisons with Alternative Analyses}

\begin{figure}[h!]
\includegraphics[scale=0.42]{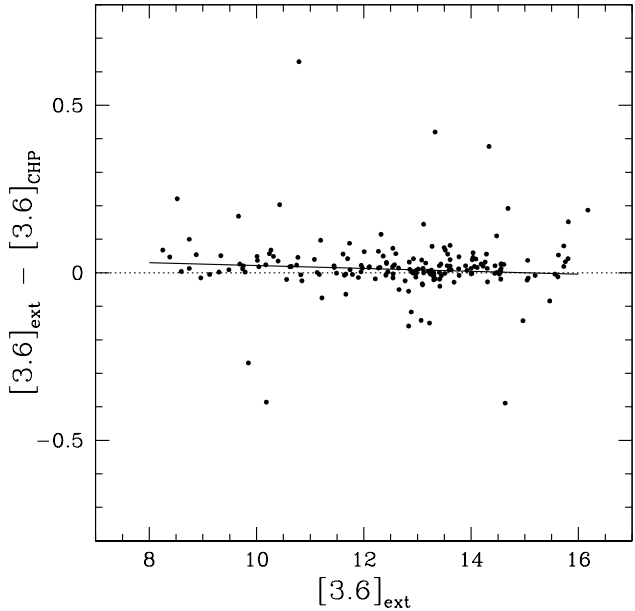}
\caption{Comparison of Archangel exponential disk extrapolated magnitudes $[3.6]_{ext}$ with aysmptotic total magnitudes from the Carnegie Hubble Program $[3.6]_{CHP}$.  The slight tilt of the thick line best fit, and the offset of $<[3.6]_{ext} - [3.6]_{CHP}> = 0.01$ from the dotted line null value, have only $2-2.5 \sigma$ significance. Both the tilt and the offset have been computed rejecting 6 deviant points.} 
\label{chp}
\end{figure}

Our Archangel analysis procedures can be compared with alternative reductions of Spitzer observations.  Comparisons with magnitudes found by the projects SINGS \citep{2009ApJ...703.1569M}, S4G (\citet{2010PASP..122.1397S}, private communication), and CHP (\citet{2011AJ....142..192F}, private communication) show that our Cosmic Flows project is on the same magnitude scale as all these projects. In the case of CHP we give special attention to a comparison because our two programs, CHP and CFS, have the common ambition of measuring galaxy distances. As we go forward, we want to understand to what degree the alternative photometry analyses are interchangeable.  A comparison is given between the two sources in Figure~\ref{chp}.  There is a slight tendancy for CHP values to be brighter for the largest galaxies, with essentially no difference faintward of $[3.6] = 12$.  The most likely explanation for a difference with the bright, large galaxies is small differences in the way sky values are set.  The rms scatter in the differences (6 deviant points rejected) is $\pm 0.052$ which, if attributed equally, implies an uncertainty in an individual measurement of $\pm 0.037$ mag for each source. \\

Comparisons with other projects give comparable results.  Typical zero point differences are $\pm 0.01$ and rms uncertainties are $\pm 0.04-0.05$ mag.  
See a summary of comparisons with other major programs in Table~\ref{table2}.  These results provide an estimate of the internal errors of alternate fitting procedures with the same data.
We recall that our two measures of magnitude agree at the level of 0.01 with scatter $\pm 0.02$. \\

\begin{table}[h!]
\begin{center}
\begin{tabular}{|c|c|c|c|c|}
\hline
1&2&3&4&5\\
\hline
Program & N & Range & $m_{cfs}-m_{other}$  & rms \\
\hline
 SINGS & 12 & 8-10 & -0.02 & 0.07 \\
 Jarrett & 5 & 9-11 & -0.01 & 0.03 \\
 S4G & 29 & 9-16 & -0.01 & 0.04 \\
 CHP & 171 & 8-16 & 0.01 & 0.05 \\
\hline
\end{tabular}
\caption{Comparisons between CFS magnitudes and other magnitudes: (1) program name, (2) number of galaxies compared, (3) range of magnitudes, (4) difference CFS magnitude $-$ other magnitude, and (5) scatter.}
\label{table2}
\end{center}
\end{table}

In summary of errors, the dominant contributions are sky settings (0.04 mag), flux calibration (0.02 mag), and extinction (0.02 mag), leading to total uncertainties in magnitudes of $\sim 0.05$ mag.   The great interest with the {\it Cosmic Flows} program is to use TFR to measure distances to galaxies and the typical scatter in TFR is 0.4 mag, 20\% in distance.  With errors on photometry after corrections held to 0.05 mag the contribution to the distance error budget from photometry is minor.


\section{Conclusions}

Our photometric procedures for the semi-automated analysis of Spitzer IRAC channel 1 data at $3.6 \mu$m have been described.  The galaxy surface photometry is carried out with the Archangel software \citep{2007astro.ph..3646S,2012PASA...29..174S} adapted for Spitzer data input.  Already material is available for some 3000 galaxies from the Spitzer Heritage Archive and our {\it Cosmic Flows with Spitzer} program will supply information for an additional 1274 galaxies. The 235 galaxies analyzed in the course of this paper will be used in a subsequent paper for the calibration of the mid-infrared Tully-Fisher relation.\\
	 
	The final goal of our project is to measure distances, hence map peculiar velocities, across the local universe within $10,000$~km s$^{-1}$ using the correlation between galaxy luminosities and their rotation rates.  We have demonstrated the ability to use {\it Spitzer Space Telescope} mid-IR data to perform surface photometry with a relatively high accuracy.   No correlation is found between magnitude uncertainties and other important galaxy parameters such as inclination, apparent area, or semi-major axis.  We conclude that, after all corrections, uncertainties on magnitudes are of the order $\pm 0.05$ for the regular spiral galaxies at the heart of our project.  These uncertainties are small compared with the overall scatter in the TFR. Low surface brightness galaxies or very irregular ones require special attention but these classes of galaxies are not of principal interest to us. \\

For our purposes, the advantages of mid-infrared photometry from space include minimization of both galactic and internal obscuration issues, very low backgrounds, and source fluxes dominated by old stellar populations that are good representatives of the baryonic mass.  The most outstanding advantage, though, is the integrity and consistency of the photometry in all quadrants of the sky.  The Extragalactic Distance Database, EDD, contains HI profile information that provides useful line widths for over 11,000 galaxies.  Ongoing Spitzer observations are providing the complementary photometric information required for a dense, detailed map of structure and motions in the near part of the Universe.\\
	
\bigskip

Acknowledgements. Comparisons between alternative analysis procedures have been facilitated by data made available by Tom Jarrett, Mark Seibert representing CHP, the {\it Carnegie Hubble Program}, and Kartik Sheth, representing S4G, the {\it Spitzer Survey of Stellar Structure in Galaxies}.  Thanks to James Shombert for the development and support of the Archangel photometry package.   The referee Michael Pohlen encouraged us to write the overview of {\it Cosmic Flows} which constitutes the Appendix and was otherwise very helpful.
NASA through the Spitzer Science Center provides support for CFS, {\it Cosmic Flows with Spitzer},  cycle 8 program 80072.  In addition to the authors, CFS co-investigators are Wendy Freedman, Tom Jarrett, Barry Madore, Eric Persson, Mark Seibert, and Ed Shaya.  
RBT receives support for aspects of this program from the US National Science Foundation with award AST-0908846.

\bigskip\noindent{\bf Appendix: Cosmic Flows Program Overview}\\

{\it Cosmic Flows} may have as many arms as an octopus.  At its core is a collaboration between Courtois and Tully to obtain accurate distances to galaxies.   A major part of the program involves exploitation of the TFR.   Activities in this regard began with the accumulation of HI profiles for the necessary kinematic information within the {\it Cosmic Flows Large Program} using the US National Radio Astronomy Green Bank Telescope \citep{2009AJ....138.1938C, 2011MNRAS.414.2005C} and the accumulation of optical photometry for the necessary magnitude and inclination information using the University of Hawaii 2.24m telescope \citep{2011MNRAS.415.1935C}.   The present extension embarks on complementing the optical photometry with mid-infrared photometry.  The current paper describes analysis procedures developed within the core program but {\it Cosmic Flows with Spitzer} embraces the larger team identified in the acknowledgement.  The next paper in this series will include most of the CFS team in a discussion of the TFR calibration with Spitzer photometry.\\

Near, intermediate, and far TFR samples in the {\it Cosmic Flows} program were described by \citet{2011MNRAS.414.2005C}.  The `near' sample is intended to achieve dense coverage of a volume extending to 3300~\kms\ with inclusion of all galaxies typed later than Sa that are brighter than $M_K =-21$, inclined greater than $45^{\circ}$, and not obscured, disrupted, or confused.  The `intermediate' sample is drawn from flux and color limits applied to an Infrared Astronomical Satellite redshift survey \citep{2000MNRAS.317...55S}.  The flux limit at $60~\mu$m is 0.6 Jy, the color criterion to separate normal spirals from active nuclei is a ratio of $100~\mu$m to $60~\mu$m flux greater than one, there is a velocity cutoff at 6000~\kms, and there is the same inclination restriction as with the near sample.  By contrast, the `far' sample is restricted to extreme edge-on systems drawn from Flat Galaxy catalogues \citep{1999BSAO...47....5K, 2004BSAO...57....5M}.  Candidates in the sample that lie at declinations accessible to Arecibo Telescope have velocities extending to 15,000~\kms.  These are our well defined samples.  In addition we derive distances to all other suitably observed galaxies.  Generally the information for the additional systems comes from archives.  In all, presently good data are available for about 7500 appropriate galaxies.\\

A quite separate and active component of {\it Cosmic Flows} is a program with {\it Hubble Space Telescope} to obtain Tip of the Red Giant Branch distances to nearby, spatially resolved galaxies \citep{2006AJ....132.2729M, 2007ApJ...661..815R, 2009AJ....138..332J}. Exquisite distances (5\% accuracy) are available for approaching 300 galaxies within $\sim 10$~Mpc.\\

Distances for {\it Cosmic Flows} encompass measures by other methodologies discussed in the literature.  Foremost among these are Cepheid Period-Luminosity Relation, Surface Brightness Fluctuation, Fundamental Plane, and Supernova Ia procedures.  The diverse material is drawn together in EDD, the Extragalactic Distance Database\footnote{http://edd.ifa.hawaii.edu} \citep{2009AJ....138..323T}.  EDD goes beyond the compilation of catalogs relevant to extragalactic distances to include redshift catalogs, that with various levels of completion describe the distribution of galaxies in the local universe, and group catalogs, that help identify entities where averaging over velocities or distances is reasonable.  The first assembly of distances in this program \citep{2008ApJ...676..184T} has now been given the name {\it Cosmicflows-1}.  A core team is now involved in the assembly of {\it Cosmicflows-2} \citep{2012ApJ...749...78T, 2012ApJ...749..174C}.\\

The holy grail of {\it Cosmic Flows} is the use of distances to determine peculiar velocities and, subsequently, mass fluctuations.  Peculiar velocities are departures from the cosmic mean expansion and it is assumed that they arise due to density irregularities.  Two regimes require separate attention.  The high density environments in and around collapsed halos are at the extreme of non-linear dynamics.  Within the collaboration we have developed Numerical Action Methods that provide an optimal description of the distribution of mass affecting galaxies on curved orbits on first approach to an attractor \citep{1995ApJ...454...15S, 2001ApJ...554..104P, 2011arXiv1105.5596P}.  The other extreme is the regime of linear dynamics.  A procedure we have used that is appropriate with redshift data sets of $10^5$ or more objects is based on the action principle \citep{2010ApJ...709..483L}.  However the methodology that most interests us starts with Wiener filtering of the peculiar velocity field resulting in descriptions of the density field independent of information provided by redshift surveys \citep{1995ApJ...449..446Z, 2012ApJ...744...43C}. The current density field can be mapped back to initial conditions that are then the starting point for constrained simulations that attempt to approximate the observed universe with a computer model \citep{2003ApJ...596...19K, 2010arXiv1005.2687G, 2012arXiv1205.4627C}.

\clearpage

\bibliographystyle{apj}

\bibliography{bibli1}

\end{document}